\documentclass[amsmath,amssymb, reprint,numerical,tightenlines]{revtex4-2}

\usepackage{graphicx}
\usepackage{dcolumn}
\usepackage{bm}
\usepackage{graphics}
\usepackage[utf8]{inputenc}
\usepackage[T1]{fontenc}
\usepackage{amsmath,booktabs,braket,cleveref,mathptmx,xcolor,siunitx,paralist}
\usepackage[version=4]{mhchem} 
 \usepackage[normalem]{ulem}
\usepackage{array}
\newcommand{\Abinitio}{\emph{Ab initio}}
\newcommand{\abinitio}{\emph{ab initio}}
\newcommand{\Duo}{{\sc Duo}}
\newcommand{\Level}{{\sc Level}}

\newcommand{\Marvel}{{\sc Marvel}}
\newcommand{\cm}{cm$~{}^{-1}$}

\def\a0{{$a_{\rm 0}$}}

\newcommand{\alert}[1]{\textcolor{red}{ #1}}

\newcommand{\mc}{\multicolumn}
\newcolumntype{H}{>{\setbox0=\hbox\bgroup}c<{\egroup}@{}}

\graphicspath{{Figs/}}

\begin{document}
\title{Molecular diatomic spectroscopy data}

\author{Laura K. McKemmish}
\affiliation{School of Chemistry, University of New South Wales, 2052 Sydney}
 \email{l.mckemmish@unsw.edu.au}


\begin{abstract} 
Accurate and comprehensive diatomic molecular spectroscopic data have long been vital in a wide variety of applications for measuring and monitoring  astrophysical, industrial and other gaseous environments. These data are also used extensively for benchmarking quantum chemistry and applications from quantum computers, ultracold chemistry and the search for physics beyond the standard model. Useful data can be highly detailed like line lists or summative like molecular constants, and obtained from theory, experiment or a combination. 

There are plentiful (though not yet sufficient) data available, but these data are often scattered.
For example, molecular constants have not been compiled since 1979 despite the existing compilation still being cited more than 200 times annually. Further, the data are interconnected but updates in one type of data are not yet routinely applied to update interconnected data: in particular, new experimental and \abinitio\ data are not routinely unified with other data on the molecule.

This paper provide information and strategies to strengthen the connection between data producers (e.g. \abinitio\ electronic structure theorists and experimental spectroscopists), data modellers (e.g. line list creators and others who connect data on one aspect of the molecule to the full energetic and spectroscopic description) and data users (astronomers, chemical physicists etc). All major data types are described including their source, use, compilation and interconnectivity. Explicit advice is provided for theoretical and experimental data producers, data modellers and data users to facilitate optimal use of new data with appropriate attribution.   

\end{abstract}


\maketitle





\emph{This is a pre-print of an article published in WIRES Computational Molecular Science (prior to peer review process). The final authenticated version is available online at: https://doi.org/10.1002/wcms.1520}

\section{Introduction}

Over the decades, there have been thousands of theoretical and computational studies of the high resolution spectroscopy of hundreds of diatomic molecules \cite{HH,01BeMc,Data2020.ExoMol}. These data are used across the sciences including monitoring pollutants like CO, NO, SO in Earth's atmosphere \cite{hitran2016}, understanding astrophysical objects \cite{fortenberry2017quantum} like planets \cite{konopacky2013detection} and stars \cite{tsuji1986molecules} using molecules like TiO \cite{piette2020assessing}, assessing the accuracy of theoretical methodologies \cite{o2005benchmark,styszynski2010we}, creating and manipulating ultracold molecules,  assessing the suitability of a molecule as a probe of variation of fundamental constants \cite{flambaum2007enhanced,19SSyMoCu.mpme}, for laser cooling \cite{10ShBaDe.lasercool} and for quantum computing \cite{demille2002quantum}.

The type of molecular spectroscopic data produced varies considerably but includes:  
\begin{compactitem}
    \item assigned experimental transitions (spectral lines);
    \item energy levels;
    \item spectroscopic networks with self-consistent assigned transitions and energy levels with uncertainties;
    \item model Hamiltonian parameters, especially band and equilibrium constants; 
    \item line lists, i.e. sets of energy levels plus intensities of transitions between these levels;
    \item potential energy, spin-orbit and other energetic coupling curves;
    \item dipole and transition dipole moment curves;
    \item cross-sections of absorption vs frequency. 
\end{compactitem} The data can be fully \abinitio\, experimental or modelled from a combination of both.  

As a simplification, here I categorise researchers as data producers (notably \abinitio{} electronic structure theorists and experimental spectroscopists), data modellers (e.g. line list and spectroscopic network creators) and data users (e.g. astronomers, chemical physicists). Data producers generally obtain data on one aspect of a molecule, e.g. assigned transitions for a spectral band or dipole moment curves for the three lowest lying electronic states. Data modellers combine  raw data into a format suitable for use in applications. For example, modellers may extrapolate from a small number of measured spectral lines to predict all transitions within frequency and intensity thresholds. Data modellers may also consolidate data from multiple individual experiments, for example, a harmonised set of molecular constants. The motivations and goals of data users are very diverse, with the group unified by the desire for easy and quick access to high-quality data with clearly communicated strengths and limitations.

Diatomic spectroscopic data are plentiful, yet contrary to expectations, existing data are not always sufficient for applications. For example, detection of molecules in exoplanets using cross-correlation of high-resolution ground-based measurements requires molecular line positions to be known to sub \cm{} accuracy; for TiO this necessitated the creation of a new line list \cite{Toto}. As another example, europium diatomics are of potential interest in measuring variation in fundamental constants but the properties of their electronic states are not generally known \cite{Victor}.

Problematically, available data on a single molecule are usually quite dispersed and it is an unexpectedly formidable task to collate and consolidate all available data. In the age of big data and machine learning, this is particularly concerning as the quality of any data science and/or artificial intelligence analysis depends first on the availability of well curated data sets. The difficulty of finding appropriate data means that many authors understandably often refer back to the last major collation of data on diatomic molecules; the famous Huber-Herzberg dataset of equilibrium molecular constants \cite{HH} (henceforth HH), published in 1979 after a decade of compilation.  This data synthesis was hugely successful, with more than 9000 citations (depending on source) including at least 200 citations annually in the 2010s. 



Though HH was a exemplery effort of its time, the data was provided as tables in a book. In the 21st century, researchers expect databases to be digitised, accessible and queryable both graphically through a website and programmatically through API. Such databases are certainly available for some types of spectroscopic data and some molecules (see \Cref{tab:databases}). However, this is not universal and some types of data, particularly \abinitio{} data, do not appear to be compiled at all, an issue I highlight as a future opportunity. 

A more challenging issue for our community is ensuring our databases are easily and regularly updated, with version control and thorough documentation of both initial data production and subsequent modifications. Often, data is interconnected; for examples, line lists rely on assigned experimental transitions and \abinitio{} dipole moment curves. The technology and database systems are now in place for many (but not all) types of spectroscopic data to facilitate these updates and streamline dissemination and utilisation of new data by users and in other types of data. However, these update mechanisms are rarely utilised by new data creators. 

Some barriers restricting effective data production, consolidation and frequent updating include: (1) the diversity of sub-disciplines involved in producing, consolidating and utilising these data, (2) the variety of data types and complexity of their interconnections, (3) insufficient communication and understanding between sub-disciplines on the most important data requirements, (4) technological issues with unfamiliar software and data formats, and (5) release of raw data and the inclusion of individual data into large databases, or individual databases into a centralised database, can (or at least is perceived to) reduce the citations to individual papers. 

To address these barriers, this focus review has two main parts. In Section II, I aim to enhance interdisciplinary understanding by concisely describing all the key types of spectroscopic data for diatomic molecules, identifying their interconnectivity, method of production, common use and major compilations. Then in Section III, I provide specific  recommendations for four key groups - experimental spectroscopists, theoretical quantum chemists, data modellers and data users - to ensure new data is effectively and easily utilised as well as appropriately attributed.

\section{\label{sec:overview}Diatomic Molecular Spectroscopic Data}

\subsection{Overview}
This section describes major types of diatomic molecule data, their source and major compilations. 

\begin{figure*}
    \centering
    \includegraphics[width=\textwidth]{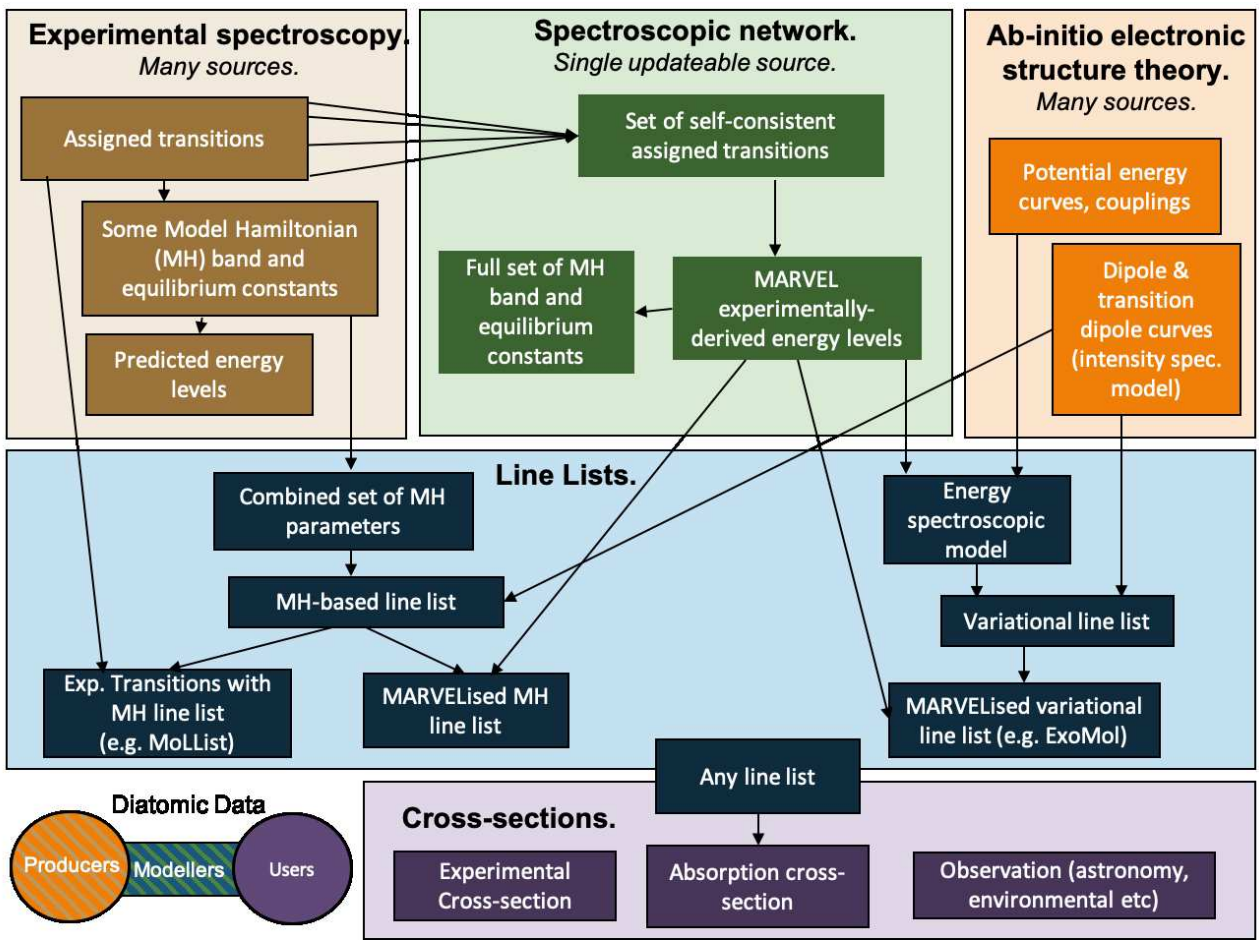}
    \caption{\label{fig:AllMolecularData} Connection between common types of diatomic molecular spectroscopic data.}
\end{figure*}

\Cref{fig:AllMolecularData} provides a flowchart showing the interconnectivity of data types. Data is divided into five main categories: experimental spectroscopy, \abinitio{} electronic structure theory, consolidated experimental spectroscopic network (usually from a \Marvel{} analysis), line list construction and cross-sections. 

\Cref{tab:databases} identifies major compilations of each data type; these are discussed further in the relevant sub-section.

\begin{table*}
\renewcommand{\arraystretch}{1.2}
\footnotesize
\caption{\label{tab:databases} Major data compilations of the spectroscopy of diatomic molecules. Data types abbreviations are: SpecN = Spectroscopic network with self-consistent set of assigned transitions and associated energy levels, EqConst = Equilibrium constants, LL = line list (i.e. transition positions and intensities), VLL = Variational LL, MHLL = Model Hamiltonian line list,  SM = energy and intensity spectroscopic model (potential energy curves, dipole and transition dipole curves etc), FCF = Franck-Condon factors, Lit = compiled literature references.  Number of molecules (\# Mols) and isotopologues (\# Isos) is for diatomics only.}
\begin{tabular}{>{\textbf}p{3cm}>{\centering}p{2cm}HHp{6cm}HcHp{5.7cm}}
\toprule
Source  & Data Type & Data Type & Primary Data & Strengths & Coverage & \# Mols & \# Isos & Molecules \\
     \midrule
\mc{5}{l}{\textbf{Spectroscopic Networks}}  \\
\Marvel \cite{Marvel,Marvel2,Marvelonline}& SpecN, Lit & All Assigned Transitions, Experimentally-derived Energy Levels & Exp & Centralised and self-consistent compilation of all experimental transitions data, uncertainties provided.  &  & 10 & \alert{xxx} & AlH, BeH, \ce{C2}, CN, CP, NH, NO, \ce{O2}, TiO, ZrO \\
\vspace{-0.5em} \\

\mc{5}{l}{\textbf{Molecular Constants}} \\
Huber-Herzberg 1979 \cite{HH} & EqConst &  Equilibrium constants & $>$ 7500 exp. & Extensiveness of molecular coverage, now available online via NIST Chem Webbook. & All diatomics  & >100 &  & Many \\
Diatomic Molecular Spectroscopic Database (DMSD) \cite{DMSD} & EqConst, FCF & & &  Modern queryable database and user data contributions facilitated and encouraged; initially populated with some HH data & & 173 & & Many; 132, 35 and 6 molecules with $\Sigma$, $\Pi$ and $\Delta$ ground states respectively \\
\vspace{-0.5em} \\

\mc{5}{l}{\textbf{Rovibronic Line lists} } \\

HITRAN \cite{hitran2016,HITRANonline} & LL & Line list & \alert{Usu. exp.} & Designed for Earth atmosphere applications with verified data of very high accuracy, with uncertainties and various line broadening parameters. & Earth-atmosphere relevant molecules  &  13 & 38 &  CO, CS, ClO, HBr,  \ce{HF}, \ce{HI}, \ce{N2}, NO, \ce{O2}, \ce{OH}, \ce{NO+} \\
HITEMP \cite{rothman2010hitemp} & LL & Line list & Usu. exp.  & Similar to HITRAN but application to higher temperatures. & Higher temperature  &  3 & 12 &  CO, NO, OH \\
ExoMol-main \cite{exomol1,exomol2,exomol3,Data2016.ExoMol,Data2020.ExoMol,Diatomics.ExoMol} & VLL, EL, SM & Variational line lists with energies & Exp, Theory & Completeness of each line list and validity at high temperature. Describes isotopologues easily and well. & Exoplanet-relevant, various & 26 & & AlH, AlO, BeH,  CaH,  CaO, \ce{C2}, CS, HS, KCl, MgH,  MgO, NaH, NaCl, NO, NS,  PN, PH, PO, PS,  ScH, SH, SiH, SiS, SiO,   TiO,  VO\\
MoLLIST \cite{MoLLIST} & MHLL &  Perturbative line lists & Exp, Theory  & Experimental basis, complementary coverage. &  & 20  & & AlCl, AlF, CaH, CaF, \ce{C2}, CH, CN,  CP, CrH, FeH, KF, LiCl, LiF, MgF, MgH, NH, NaF, OH, \ce{OH+}, TiH\\

ExoMol-collated \cite{Data2020.ExoMol,exomol.empirical}  &  LL & & &  Central repository in single format for various rovibrational \& rovibronic line lists, esp. ExoMol-main and MoLLIST & & 52 & 147  & Non-MoLLIST/ExoMol-main: \ce{H2}, \ce{HF}, \ce{HCl}, \ce{N2}, \ce{SiO}, \ce{SiO}, \ce{ScH}, \ce{LiH}, \ce{LiH+}, \ce{CO}, \ce{HeH+}, \ce{HD+}, \ce{HD}, \ce{YO}   \\

Kurucz \cite{Kurucz2011,Kurucz2017} & LL & & &  Breadth of included molecules and consistency of data compilation over time. & & 45 & & Notable: BH, BO, \ce{CH+}, \ce{CN+}, CO, \ce{CO+}, CrH, CS, FeH, \ce{H2}, KH, LaO, LiH, MgH, \ce{MgH+}, MgO, NH, \ce{NH+}, NO, NS, NaH, \ce{O2}, ScO, SiC, \ce{SiH+}, \ce{SiO+},SiS, TiH, YO, ZrO \\
GEISA 2015 \cite{GEISA} & LL & & & & Gestion et Etude des Informations Spectroscopiques Atmosphériques (GEISA)  & 11 & 25 & ClO,  CO, HBr,  HCl, HF, HI,   \ce{N2},NO,  \ce{NO+}, \ce{O2}, OH  \\
Plez \cite{plez,plez1998new}  & LL & & & & & 17 & & AlH, \ce{C2}, CaH, CH, CN, CO, CrH, FeH, MgH, NH, OH, SiH, SO, SiS, TiO, VO, ZrO \\
VALD \cite{VALD,VAMDC_VALDupdate} & LL & & &  Vienna Atomic Line Data Base with some molecular line lists. & & 8 & 26  & \ce{C2}, CH, CN, CO, MgH, SiH, OH, TiO  \\
    \vspace{-0.5em} \\

\mc{5}{l}{\textbf{Microwave and Millimeter-focused Line lists}}  \\
Splatalogue \cite{Splatalogue} & LL & & & Single portal for all LL data below &  & >100 &  &  \\ 
CDMS \cite{CDMS1,CDMS2,CDMS3} &LL &  & &  Cologne Database for Molecular Spectroscopy  & & 84 & 181 &  Many \\
JPL  \cite{JPL,JPLwebsite}  & LL &  & &  Jet Propulsion Laboratory Millimeter and Submillimeter Spectral Line Catalog & & 54 & 91 & Many  \\
Lovas/NIST \cite{Lovas} & LL & & & & & 25 & 55 & Many \\

ToyaMA \cite{ToyaMA} & LL & & & & & 3 & 3 & NF, NCl, OD$^-$ \\

\vspace{-0.5em} \\
\mc{5}{l}{\textbf{Compiled Literature}} \\
DiRef \cite{01BeMc} & Lit & & & More than 30,000 references  mostly from 1974-2000 based largely on the Berkeley newsletters by Phillips, David and Eakin \cite{Berkeley}. & & >100  \\
ExoMol-bibliography  &  Lit & & &  Collated references on particular molecules. & & 58 &   &    \\
     \bottomrule
\end{tabular}
\end{table*}
\renewcommand{\arraystretch}{1}




\subsection{Assigned Experimental Transitions}

\textbf{Definition:} An assigned experimental transition consists of a transition frequency between an initial and final quantum state described with approximate quantum numbers and ideally an uncertainty.

\textbf{Source:} Experimental spectroscopy papers usually provide tabulations of assigned rovibronic spectral transitions for one or more spectral bands for one or more molecules. 

\textbf{Discussion:} 
Assigned spectral transitions are the observable evidence of the molecule's energetic and spectroscopic properties. 

\textbf{Data Compilation:} Beyond individual papers (the main source of data), today this data is generally collated as consolidated experimental spectroscopic networks (see below). Previously, this type of data was usually collated as experimental line lists, in this format lacking self-consistency and often attribution to individual data sources.

\subsection{Rovibronic Energy Levels}

\textbf{Definition:} Energy of rovibronic states (usually relative to the lowest energy quantum state in the molecule and ideally with uncertainties) with associated approximate quantum numbers describing the state.

\textbf{Source:} The most accurate set of energies is obtained by inverting the set of self-consistent \Marvel{} transition frequencies to obtain experimentally-derived \Marvel{} energy levels. Rovibronic energy levels can also be calculated from model Hamiltonian band or equilibrium constants, or by solving the nuclear motion Schrodinger equation for a given energy spectroscopic model. 

\textbf{Common use:} 
\begin{compactitem}
    \item to enable fitting of the energy spectroscopic model;
    \item to replace modelled energy levels in a line list and thus enable high precision frequency predictions. 
\end{compactitem} 




\subsection{\label{subsec:SN}Consolidated Experimental Spectroscopic Network}

\textbf{Definition:} A large set of self-consistent assigned experimental transitions with uncertainties, and the associated inverted experimentally-derived energy levels with uncertainties.

\textbf{Production:}
Spectroscopic networks are usually produced through what has become known as a \Marvel{} compilation and analysis (after the key enabling program). This involves: (1) a thorough literature review, (2) digitisation and consolidation of all available assigned transitions into a single format, (3) using the \Marvel{} software program, process data within first individual sources then multiple sources to produce a self-consistent spectroscopic network by increasing uncertainties and removing transitions as required to ensure self-consistency of the whole spectroscopic network.

\textbf{Compilation:} Centralised to the \Marvel{} online website with modern accessability and database standards. 

\textbf{Updates:} 
New experimental data should be immediately added to available \Marvel\ datasets, providing the dual benefits of cross-validation and ensuring the new data can be used easily to update line lists. 
The availability of \Marvel{}online substantially reduces the barrier to this task, which can be done through a graphical user interface. However,  training with an experienced user can be useful. 

Alternatively, well-formatted plain text data as supplementary information, ideally in the molecule's \Marvel{} format, and an email to the original compilation authors will generally enable new data to be added. 

\subsection{Model Hamiltonian Parameters (e.g. Band and Equilibrium Constants)}
\label{subsec:MH}
\textbf{Definition:} The set of parameters that are used to approximate the energy of quantum states in a molecule based on each state's quantum numbers. 

\textbf{Discussion:} Reasonable functional forms for the dependency of diatomic energy levels (and thus transition frequencies) on molecular quantum numbers can be obtained by approximating the full Hamiltonian of the system, i.e. using a model Hamiltonian (e.g. a rigid rotor or Morse oscillator) \cite{brown2003rotational,lefebvre2004spectra}.  These expressions have parameters that can be adjusted to maximise the fit between the predicted and experimental transition frequencies. 

\textbf{Components:} 
The most common parameters are the band constants ($T_v$, $B_v$, $D_v$ etc) - obtained when fitting rotational levels within a single vibrational state of a single electronic state - and equilibrium constants ($T_e$, $\omega_e$, $\omega_e\chi_e$, $B_e$, $D_e$ etc) - obtained when fitting rotational levels from multiple vibrational levels of a single electronic state. For states with non-zero angular momentum and spin, the spin-orbit coupling, $A$, is also vital. 

Higher order couplings, such as the spin-spin and spin-rotation couplings, and off-diagonal couplings between electronic states are important for high accuracy applications.

\textbf{Software:} PGopher \cite{Pgopher} is the community standard program for fitting observed spectral transitions to model Hamiltonians. 


\textbf{Advantages: } 
Good summary of the key features of a molecule's electronic structure, sufficient for many applications. Can be used to calculate partition functions \cite{barklem2016partition}.
 
 \textbf{Limitations: } 
Molecular constants are problematic when there is strong coupling between electronic states that lead to perturbations in quantum energy levels \cite{lefebvre2012perturbations}. 

Energy levels and spectra predicted from molecular constants do not extrapolate beyond experimental data as reliably as variational approaches based on fitted energy spectroscopic models. 

\textbf{Sources:} 
Many papers report equilibrium and band constants only for their data. It is less common to find a synthesis of all the molecular constants for a given molecule.

\textbf{Major Compilations:}
Currently, the best compilation is still HH from 1979. The easiest way to access most of this data is the Diatomic Molecular Spectroscopy  Database (DMSD) \cite{DMSD}, which adheres to modern database standards. More importantly, DMSD is updateable by the community; inclusion of molecular constants post-1979 into DMSD is a key  priority in diatomic data collation. 

Band constants are not centrally compiled, yet are preferred to equilibrium constants for high accuracy applications. The easiest pathway forward is probably the centralised collation of PGopher input files, with appropriate author attributions. This data could then be processed into a SQL database and made accessible online via, for example, the DMSD website.

\subsection{Line Lists}
\textbf{Definition:} A line list contains the most comprehensive model of a molecule's spectroscopy, containing energy levels, transition frequencies and transition intensities. 

\textbf{Use:} Line lists are most frequently used to predict absorption cross-sections and as input to sophisticated (e.g. radiative transfer) models of gaseous environments. Line lists can also be used \cite{exocross} to compute state lifetimes and molecular thermodynamics properties such as partition functions, specific heat and cooling functions.

\textbf{Types:}
\Cref{fig:AllMolecularData} identifies five produced types of line lists that are characterised by the 
\begin{compactitem}
    \item the source of the interpolated and extrapolated energy levels and transitions - model Hamiltonians (MH) or energy spectrosocopic model (ESM) - these are sometimes called perturbative and variational line lists respectively;
    \item whether experimentally-derived \Marvel\ energy levels and/or experimental transitions are used to replace modelled energy levels. 
\end{compactitem}

A potential superior type of line list, not yet created, would use \Marvel{} energies where available, MH energies where reliable and ESM energies elsewhere.

\textbf{Software:} \Level{} \cite{LEVEL} has long been the standard program for computing diatomic line lists. However, its limited ability to describe multiple coupled electronic states means that it is being replaced by the new program \Duo{} \cite{Duo} which treats this coupling systematically. 
Both programs solve the nuclear motion Schrodinger equation to convert an ESM to a set of energy levels, then combine the rovibronic wavefunction data with the intensity spectroscopy model to produce a final line list. 


\textbf{Recommendations:} Line lists should be regularly updated as new sources of \abinitio\ energetic data and/or assigned experimental transitions become available. To assist this, all input files and data to a produced line list should be included as supplementary information to published papers; for example, \Level{} or \Duo{} input files, and the energy and intensity spectroscopic models. When experimental corrections are made to calculated data, the corrected and original data should be retained.

\textbf{Major Database Compilations: } 
 \Cref{tab:databases} highlights that line lists are the most frequently compiled type of diatomic data. There is considerable diversity in data from different databases as shown for three representative diatomics in \Cref{tab:btndatabase}.
 
Two centralised database compilation websites have become prominent - the ExoMol website \cite{Data2016.ExoMol,Data2020.ExoMol} focusing on broad wavelength coverage and Splatalogue \cite{Splatalogue} focusing on microwave lines - that have been very successful in enhanced data accessibility while retaining credit to the original line list creators. 

\renewcommand{\arraystretch}{1.2}

\begin{table*}[htbp!]
     \caption{Comparison of the composition of various data compilations for the same molecule. All frequencies are in \cm{}. \#Isos is number of isotopologues considered, \#elec is number of electronic states considered, ELs = number of energy levels, Trans = transitions, No = number, Freq = transition frequencies.}
    \label{tab:btndatabase}
   \centering
    \begin{tabular}{lp{3cm}p{3cm}cccrrrHr}
    \toprule
          &  Compilation & Source Main Iso & \# Isos & \# elec & ELs & \mc{3}{c}{Trans (main iso)} \\
         \cmidrule(r){7-9} 
         & &  & & & &  \mc{1}{c}{No} & Min Freq & Max Freq& intens? \\
         \midrule
         \ce{CO} & Kurucz A-X (2014) & &  1 & 2 &- & 396,946& 21,807 & 89,818 &  \\
          & Kurucz X-X (1990) & & 1 & 1 &- & 118,920 &  1,000 & 10,379 &  \\
          & JPL  (1997) & &  4 & 1 & -   & 91 & 3 &332   & Yes at 300K \\ 
          & CDMS & & 6 & 1 &-  & 182 & 3 & 345 &  \\
          & HITRAN (2016) & & 6 & 1 &- & 1,344 & 3 & 14,477 & Yes \\
          & HITEMP (2010) & & 6 & 1 &- &  ~18,939 & 2 & 9,274 & Yes \\
          & HITEMP (2019), Plez & Li et. al., 2015 &  6 & 1 &- & 125,496 & 2 & 22,149 & Yes \\
          & Plez & Goorvitch, 1994 & 7 & 1 & - & 19,203 & &   \\
          & GEISA  &  & 6 & 1 &- & 5,908 & 3 & 8,465 & Yes \\
          \vspace{-0.5em} \\
          \ce{SiS} & \\
          & JPL & Poynter 1979  & 4 & 1 & - & 97 & 0 & 58  \\
          & CDMS & 2019 & 12 & 1 & - & 4626 & 0 & 100 \\ 
                    \vspace{-0.5em} \\
          \ce{NO} & \\
          & HITRAN (2016) & & - & 1 & - & 251,898 & 0 & 23,726 & \\
          & HITEMP (2010) & & 3 & 1 & - & 38,537 &	0 & 9,274 \\
          & HITEMP (2019) & & 3 & 1 & - & 379,064 & 0  & 26,777	 \\
          & ExoMol (2017) & NOname & 6 & 1 & 21,688 & 2,280,366 & 0 & 40,040 \\
          & GEISA & & 3 & 1 & - & 103,701 & 0 & 9,273 & \\
          & JPL & Drouin 2010 & & 1 & - & 9756 & 0 & 333 \\
          & CDMS & Müller 2015 & 6 &1  & - & 4581 & 0 & 229  \\
         \bottomrule 
    \end{tabular}
\end{table*}
\renewcommand{\arraystretch}{1}

\subsection{\Abinitio\ Energetic Data}
\label{subsec:aienergy}
\textbf{Definition:} Data produced entirely from \abinitio\ quantum chemistry electronic structure calculations using a certain model chemistry (i.e. level of theory and basis set) that enable the calculation of the energies of rovibronic quantum states in a diatomic molecule.

\textbf{Types: }
The most important types of energetic data are the potential energy curves and the diagonal and off-diagonal spin-orbit coupling curves (or at least their equilibrium values).

\textbf{Production: } The best data \cite{Diatomics.ExoMol} for multiple low-lying electronic states usually comes from multi-reference configuration interaction (MRCI) calculations based on orbitals obtained from state-selective or state-averaged Complete Active Space Self-Consistent Field (CASSCF) calculations. For the lowest lying electronic state of a given spin-symmetry with little multi-reference character, CCSD(T) with basis set extrapolation is usually the best practical choice. 

\textbf{Common use:} 
Refined or raw \abinitio\ energetic data is often a key component of an energy spectroscopic model, and thus a line list. 

Beyond line list creation, potential energy curves and the \abinitio\ coupling are essential for studies of dynamics, e.g. through molecular dynamics. 

\textbf{Compilation:} Currently very poor compilation with most potential energy curves and related energetics data in individual papers. Furthermore, many of these papers only contain figures or fitted constants, rather than the raw data. 

\Abinitio\ vibrational frequencies for diatomics for a range of model chemistries are available on the Computational Chemistry Comparison and Benchmark DataBase (CCCBDB) \cite{CCCBDB}. 



\subsection{Energy Spectroscopic Model (ESM)}
\label{subsec:ESM}
\textbf{Definition:} The set of potential energy curves and coupling curves that together can be used with a nuclear motion program such as \Level{} \cite{LEVEL} or \Duo{} \cite{Duo} to produce rovibronic energy levels for a molecule \cite{Data2016.ExoMol}. Generally, energy spectroscopic models are fitted against available spectroscopic experimental data. 

\textbf{Production:}   Both \Level{} and \Duo{} include functionality to produce a high accuracy energy spectroscopic model (ESM) by fitting a parameterised set of potential energy curves and couplings (often obtained by \abinitio\ energetics data) to assigned experimental transitions.   

\textbf{Use:}
The most common use of an ESM is to produce a line list. 

However, ESMs can be used instead of \abinitio\ energetics data in a variety of applications, for example, to simulate reaction dynamics, and will reproduce experiment better than raw \abinitio\ data. 

\textbf{Compilation Notes:} Now routinely compiled as part of variationally-produced ExoMol line lists \cite{Data2016.ExoMol,Data2020.ExoMol}. However, older line lists or line lists produced using different methodologies often do not contain this data, somewhat restricting their usefulness of their data for some applications and making future line list improvements more challenging.

\subsection{Intensity Spectroscopic Model (i.e. Dipole and Transition Dipole Moments)}

\textbf{Definition:} The set of diagonal and off-diagonal dipole moment curves that are combined with an ESM to produce the intensities of spectral transitions within a molecular line list. 

\textbf{Production:} Intensity information is almost always purely from \abinitio\ calculations using the methods described above, rather than experimental measurements. 

\textbf{Verification:}
Accurate absolute intensities from experiment are challenging. Instead, relative intensities from lab measurements and real-world data (e.g. astronomical spectra) can be used to validate models. Further, experimental  lifetimes for excited rovibronic states can directly validate the absolute intensities in the models. 

\textbf{Compilation Notes:} Intensity spectroscopic models are now routinely included as part of ExoMol line lists. 

\Abinitio\ dipole and transition moments are available in literature, usually scattered. Unfortunately, data is often exclusively in figures and thus unusable, necessitating repeated calculations.  

\textbf{Updates:} Line lists can be trivially updated with new dipole moment curves.  

Corrections based on observational data on the relative intensity of bands are possible, but must be carefully considered and documented thoroughly. 

\subsection{Cross-sections}

\textbf{Definition:} The intensity of absorption as a function of frequency or wavelength in given environmental conditions (temperature, pressure, resolution).

\textbf{Data Sources:} 
Though cross-sections are routinely catalogued for bigger molecules (e.g. by HITRAN \cite{HITRANonline} and NIST \cite{NIST}), for diatomics the underlying data such as molecular constants are more commonly reported instead, though individual papers often have cross-sections as figures. 

Modelled cross-sections can be obtained from line lists using the appropriate software (e.g. Exocross \cite{exocross} for ExoMol-formatted line lists) at a variety of temperatures and pressures. Limitations in experimental or observational resolution can be modelled using line broadening parameters.  

\textbf{Advantages:} Cross-sections are the data type most closely linked with observable data, and can be a useful visualisation aiding interpretation of the underlying spectroscopy of the model. For example, a cross-section plot is the easiest way to identify which spectral region has the strongest absorption for a given molecule and how that absorption changes with temperature and pressure.

\subsection{Compiled Literature}

\textbf{Description:} Molecule-sorted compiled lists of a significant number of references for diatomic spectroscopic data. 

\textbf{Motivation: }
There are two main motivations: (1) collate sources of spectroscopic data that have not yet been incorporated into major databases, (2) collate sources of diverse data and information beyond that categorised above that may be useful for data users - e.g. pre-dissociation and reaction rates.

Collation of all spectroscopic data on an individual diatomic can be quite time-consuming, and requires expert knowledge and skills to ensure comprehensive coverage of crucial data. Curated and updated list of literature for each molecule covering a wide variety of different data types is thus a worthwhile pursuit, admittedly one that is challenging to justify in isolation.

\textbf{Production:} 
The diatomic spectroscopic literature is quite scattered with  significant amounts of data can only be found in the original manuscripts. Keyword based searches for molecular data on typical academic search engines are often cumbersome, especially for molecules where the diatomic form is not the most common occurrence of the element combination such as titanium oxide. Commonly, searching the citation web (i.e. references and citations) of initially identified  papers has been the most effective way of collating relevant literature, but this relies on the connectivity of all relevant literature and is time consuming with existing tools. 

\textbf{Major Compilations: }  I highlight two main sources of compiled literature data, the 2001 `Diref' database by Bernath and McLeod \cite{01BeMc} - an excellent resource but now out-dated - and the ExoMol bibliography (\url{www.exomol.com}). 

I note that HH and \Marvel{} compilations both incorporate all relevant references in their data. 



\section{Ensuring new data is effectively utilised and attributed}



\subsection{Experimentalists} 
Data, most notably assigned transitions, provided by experimental spectroscopists is crucial to achieving spectroscopic accuracy for frequencies in models of diatomic spectroscopy. These data should always be provided in plain text format to facilitate use and avoid processing challenges and potential errors. 

It is optimal to minimise the lag between production of new data and its utilisation by data users; this goal can be achieved by incorporating the update of available databases alongside the publication of the spectroscopic analysis. Specifically, equilibrium constants should be added to the DMSD (noting band constants may be incorporated in future expansions). Where already compiled, new data should ideally be combined with the existing \Marvel{} spectroscopic network, with existing line lists \Marvel{}ised to incorporate the improved experimental data. Inclusion of new data in the full spectroscopic network of the molecule is also beneficial to the verification of the experimental data assignments; previous consolidations have  led to the identification and correction of calibration errors (e.g. \cite{20Mc.C2}), and could also be useful in speeding up new assignments (noting that of course previous data can be removed if the initial spectral 
assignments are rejected). 

Of course, it is disingenuous to ignore the barriers to this use of databases. 

Data producers may fear lack of attribution to their original research when their data is compiled into databases; I recommend three main remedies:
\begin{itemize}
    \item Data modellers must ensure data structures incorporate proper attribution of data sources; see data modellers recommendations. The original source is definitely included in the DMSD and \Marvel{} data. The situation is more complex with line lists as they are synthesis of all available data. The best way to ensure your data is attributed is probably to incorporate an update to the line list within your experimental paper so that users cite your update and the original paper.
    \item Increase understanding in the user community regarding the importance of data attribution; see the data user recommendations. 
    \item Use of non-traditional metrics in quantifying the impact of your research, most notably download counts, but also the citations of the molecule-specific and general database papers and specific case studies of the importance of your data in future analysis. 
\end{itemize} 

The other barrier is the technical challenges and time associated with incorporating data into databases; recent progress has substantially reduced these barriers, specifically: 
\begin{itemize}
\item Adding \textbf{equilibrium constants} to the DMSD website is trivial requiring only text entry. Complications could arise with complex Hamiltonians. 
\item To add new measured \textbf{assigned transitions to a pre-existing spectroscopic network} requires reformatting data, then running through the \Marvel{} program using the online website first by itself then in combination with pre-existing \Marvel{} data. 
If all experiments are error-free, this is a trivial step, but in practice there is normally a process of increasing uncertainties on some transitions and removing other transitions to obtain a self-consistent network. The tutorials and new features on the \Marvel{} website are extremely helpful, but additional support could be useful for new users and in difficult cases. 
\item \Marvel{} energy levels can be readily added to an \textbf{existing line list} to create an updated \Marvel{}ised line list suitable for high-resolution purposes. This step is usually trivial involving converting between data representations. Since the energy levels (not transitions) are replaced, transitions not measured can be reproduced with experimental accuracies if upper and lower energy levels are in the experimental spectroscopic network. Isotopologue line lists can also be \Marvel{}ised by assuming that the error in the spectroscopic model is approximately isotopologue-independent, and thus the main isotopologue correction factor can be adopted. 
\end{itemize}

As a final note to experimentalists, \Marvel{} publications often explicitly identify where future experimental data would be most useful.  


\subsection{Theoreticians} 
To maximise utilisation of new data, I recommend that  \begin{enumerate}
\item  dipole moments curves are calculated alongside potential energy curves (note diagonal dipoles are usually calculated automatically but not extracted while transition moment need to be specifically requested but are vital for rovibronic spectroscopy);
\item the raw data for curves (potential energy, dipole moment) are made available in the supplementary information, ideally in plain text format;
\end{enumerate} 
Updated data on experimentally undetermined electronic states, and diagonal and off-diagonal spin-orbit coupling, either equilibrium values or curves, are also very valuable. I encourage inclusion of sample input files to enhance data reproducibility. 

Evaluating the effect of new dipole moment curves for molecules with existing line lists and spectroscopic models can be trivial, and can quickly lead to updates in the recommended line lists. With \Duo\ \cite{Duo} line lists, one can simply replace existing curves in the input file, recalculate the line list and then use the ExoCross \cite{exocross} software package to produce and compare absorption cross-sections with the old and new dipole moment curves. Quantitative evidence of the effect of the changed dipole moment on the molecule's spectral properties could strongly enhance the theoretical publication. Theoreticians should discuss their results with the original line list creators; if the theoretical approach for dipole moments is clearly superior to the original model, then the newly generated line list can be adopted as the recommended version with associated citations.  
The inclusion of new potential energy curve data into line lists is more complicated and probably would need to be considered on a molecule-by-molecule basis in discussion with the line list creator. 

Finally, broadly, there is a need for a centralised, format-consistent database of diatomic \abinitio{} data, especially for machine learning applications.  


\subsection{Data Modellers}
The ease with which data producers can incorporate new data into your models, and data users can access your data including updates is central to the effectiveness of your work. To enable this, data modellers must: 
\begin{itemize}
\item provide input files and all input data;
\item  provide well-documented software programs (e.g. using journals such as Computational Physics Communication), or utilise well-established programs;
\item  enable data access both graphically via websites and programmatically through, for example, application programming interfaces (APIs) (e.g. \cite{Data2016.ExoMol,HITRANonline});
\item ensure references to original data (e.g. for \abinitio\ calculations or experimental results) are retained and cited appropriately.
\end{itemize}

Mutually beneficial relationships between new model creators and data curators must be actively sought to address the needs of both the model creators for attribution to enable  career progression and funding, and users in terms of data accessibility. Users must be strongly encouraged to cite the original source. Download counts can complement citations and demonstrate impact and should thus be provided regularly to data contributors. 

Regular small updates of individual molecule datasets, not substantial enough to warrant a new paper, should be encouraged. These intermittent updates could be documented through non-traditional publication such as arxiv or research notes, then compiled in regular database updates (already common with many major databases e.g. ExoMol and HITRAN). To support this, version control and clear data naming is essential. Unfortunately, the author knows of multiple instances where users found data inadequate for their research almost entirely because they were using out-of-date data.

\subsection{Data Users}
The most important message to data users is the following: if you want modern data, then you need to support all the researchers involved in producing this data from experimental spectroscopists, \abinitio{} quantum chemists, line list creators and so on. The easiest way to do this is through citations to individual data producers and modellers so they can quantitatively demonstrate the usefulness of their research.  Though it is easier to just cite the database, without the citations to their individual data, data producers cannot justify the time required to put their data into an easily accessible form in a database - in fact, it might be counterproductive for them to do so! 

When data is particularly critical for your research, I would encourage you to go further and send data producers and/or modellers your paper with a short summary of how you used their data. Data is often used in more diverse applications than originally anticipated, and by letting data producers know about this diversity you can enhance the strength of their next proposal to collect the data you desire. 

If you are seeking data that is not currently available or there are problems with the existing data, I encourage you to reach out. 

\section{Conclusions}

New high-resolution spectroscopic measurements and first-principles rovibronic calculations on small molecules are often motivated  by the molecule's importance in the accurate modelling and understanding  of astrophysical, environmental and atmospheric environments.  To enable the new insights provided by these measurements and calculations to be utilised by applications experts,  the data must be incorporated into centralised spectroscopic databases in a consistent, reliable, robust, and documented form. 

The links between new spectroscopic data production and the use of spectral data in scientific and engineering applications can be greatly facilitated by new data-science approaches.  This paper provides specific recommendations to experimental spectroscopists, \abinitio{} quantum chemistry theorists, data modellers and data users to take advantage of these new approaches and maximise the speed and impact of their research. Data availability as plain text and incorporated into databases is prioritised, along with an appeal to include citations to primary data sources not just consolidated databases.   

High quality updateable and queryable databases are available for many diatomic species with line lists, assigned experimental transitions and derived empirical energy levels data; expansion of these databases to other diatomics is ongoing research. The field would benefit from centralised repositories of \abinitio{} electronic structure calculation results in a plain text format. The new Diatomic Molecular Spectroscopic Database (DMSD) \cite{DMSD} incorporates many Huber-Herzberg equilibrium constants into a modern updateable database and should be embraced by the community as the path forward for this data type, and ideally expanded to the collation of molecular band constants.  

\section*{Acknowledgements} 
Thank you to all of the many people who have collaborated with me over the years on diatomic molecules to showcase the potential of these deceptively complex little systems, in particular Anna-Maree Syme, Jonathan Tennyson, Sergey Yurchenko, Maire Gorman and Lorenzo Lodi, and for comments on this manuscript draft from Juan Camilo Zapata, Anna-Maree Syme and Scott Kable.



\end{document}